\documentclass[aps,prb,twocolumn,groupedaddress,showpacs]{revtex4}

\usepackage{longtable}
\usepackage{graphicx}
\usepackage{dcolumn}
\usepackage{bm}
\usepackage{amsmath}
\usepackage[psamsfonts]{amssymb}
\usepackage{subfigure}

\newcommand{\STO}{SrTiO$_3$}
\newcommand{\LAO}{LaAlO$_3$}

\newcommand{\MR}{magnetoresistance}

\newcommand{\etal}{\emph{et al.}}

\begin{document}

\title{Phase Diagram of Micron-Size Bridges of SrTiO$_3/$LaAlO$_3$ Interface: Link Between Multiple Band Structure and Superconductivity}


\author{M. Ben Shalom}
\affiliation{Raymond and Beverly Sackler School of Physics and Astronomy, Tel-Aviv University, Tel Aviv, 69978, Israel}
\author{I. Neder}
\affiliation{Raymond and Beverly Sackler School of Physics and Astronomy, Tel-Aviv University, Tel Aviv, 69978, Israel}
\author{A. Palevski}
\affiliation{Raymond and Beverly Sackler School of Physics and Astronomy, Tel-Aviv University, Tel Aviv, 69978, Israel}
\author{Y. Dagan}
\email[]{yodagan@post.tau.ac.il} \affiliation{Raymond and Beverly Sackler School of Physics
and Astronomy, Tel-Aviv University, Tel Aviv, 69978, Israel}


\date{\today}

\begin{abstract}
The rich phase diagram of the two dimensional electron gas (2DEG) at the \STO/\LAO~ interface is probed using Hall and longitudinal resistivity. Thanks to a special bridge design we are able to tune through the superconducting transition temperature T$_c$ and to mute superconductivity by either adding or removing carriers in a gate bias range of a few volts. Hall signal measurements pinpoint the onset of population of a second mobile band right at the carrier concentration where maximum superconducting T$_c$ and critical field H$_c$ occur. These results emphasize the advantages of our design, which can be applied to many other two dimensional systems assembled on top of a dielectric substrate with high permittivity.
\end{abstract}
\maketitle
Recently, a significant progress in studying transition metal oxide interfaces has been achieved \cite{mannhart2010oxide}. It is often attributed to developments in thin film fabrication and characterization, enabling deposition of sharp interfaces on top of smooth single terminated surfaces. Due to their strongly correlated $d$ electrons, transition metal oxides exhibit a rich electronic phase diagram with engineered properties such as: insulators with large dielectric constant $\varepsilon$, high mobility systems of two dimensional electron gas (2DEG), ferroelectric, ferromagnetic and superconducting interfaces \cite{hwang2012emergent}. All these phases can be susceptible to moderate changes in carrier concentration.
\par
The interface between \STO~and \LAO~exhibits tunable superconductivity \cite{caviglia2008electric}, magnetism \cite{salman2012nature, lee2013titanium, brinkman2007magnetic} these two orders seem to coexist at the same temperature-field domain in the phase diagram \cite{bert2011direct, li2011coexistence, dikin2011coexistence} and may be affected by the strong and tunable spin-orbit interaction \cite{PhysRevLett.104.126802, caviglia2010tunable}. We have also found evidence for the existence of a multiple band structure \cite{shalom2010shubnikov, lerer2011low}. Joshua \etal~suggested that the second band becomes populated at a critical density where the transport properties such as the magnetoresistance and the anomalous Hall effect change\cite{joshua2012universal}.
\begin{figure}
\includegraphics[width=0.8\hsize]{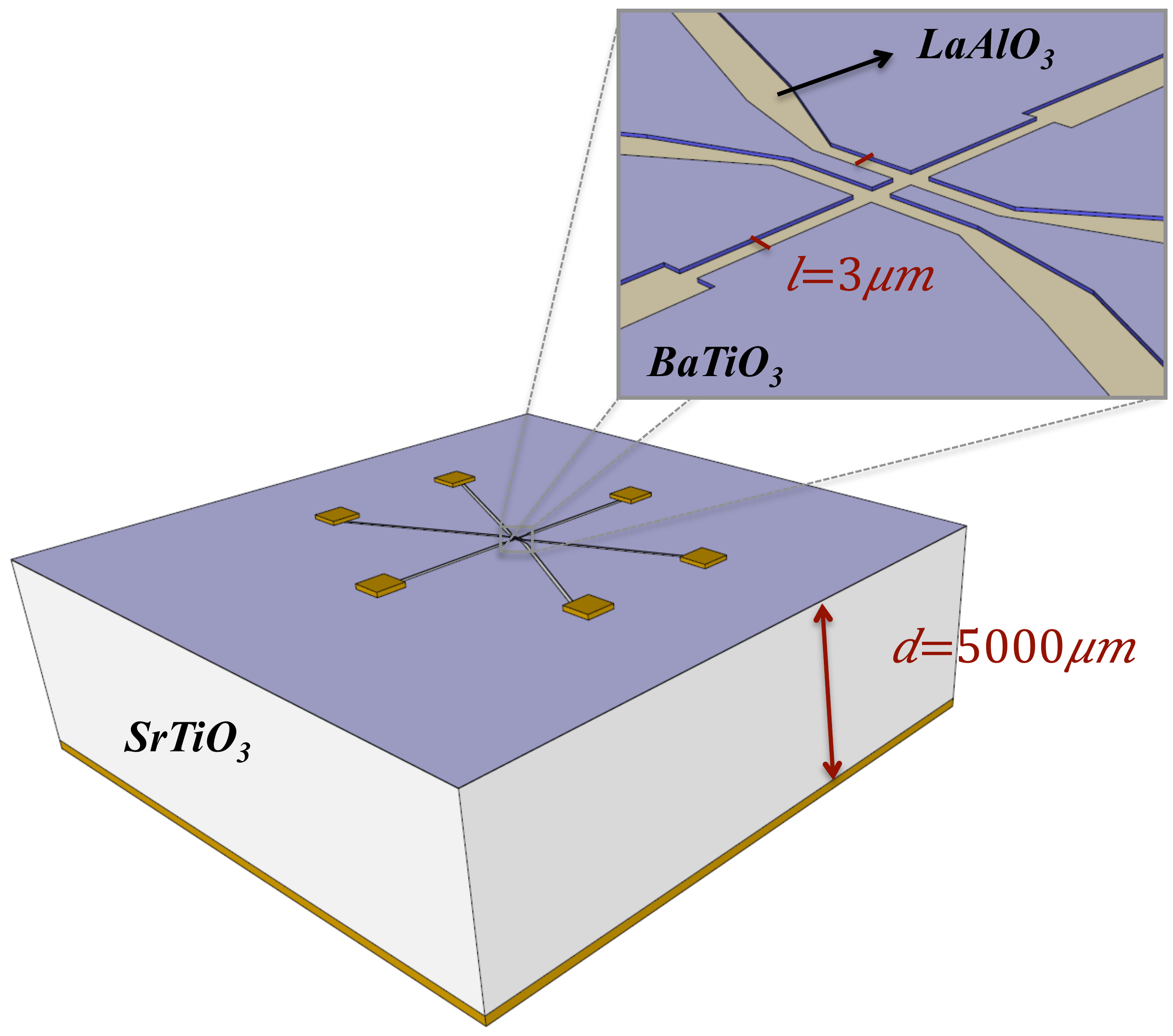}
\caption {Sketch of the mesoscopic conducting channel at the interface with the various layers needed for sample fabrication.\label{device}}
\end{figure}

\begin{figure}
 \begin{center}

  \subfigure{\hspace*{-0cm}\includegraphics[height=0.8\hsize,width=0.91\hsize]{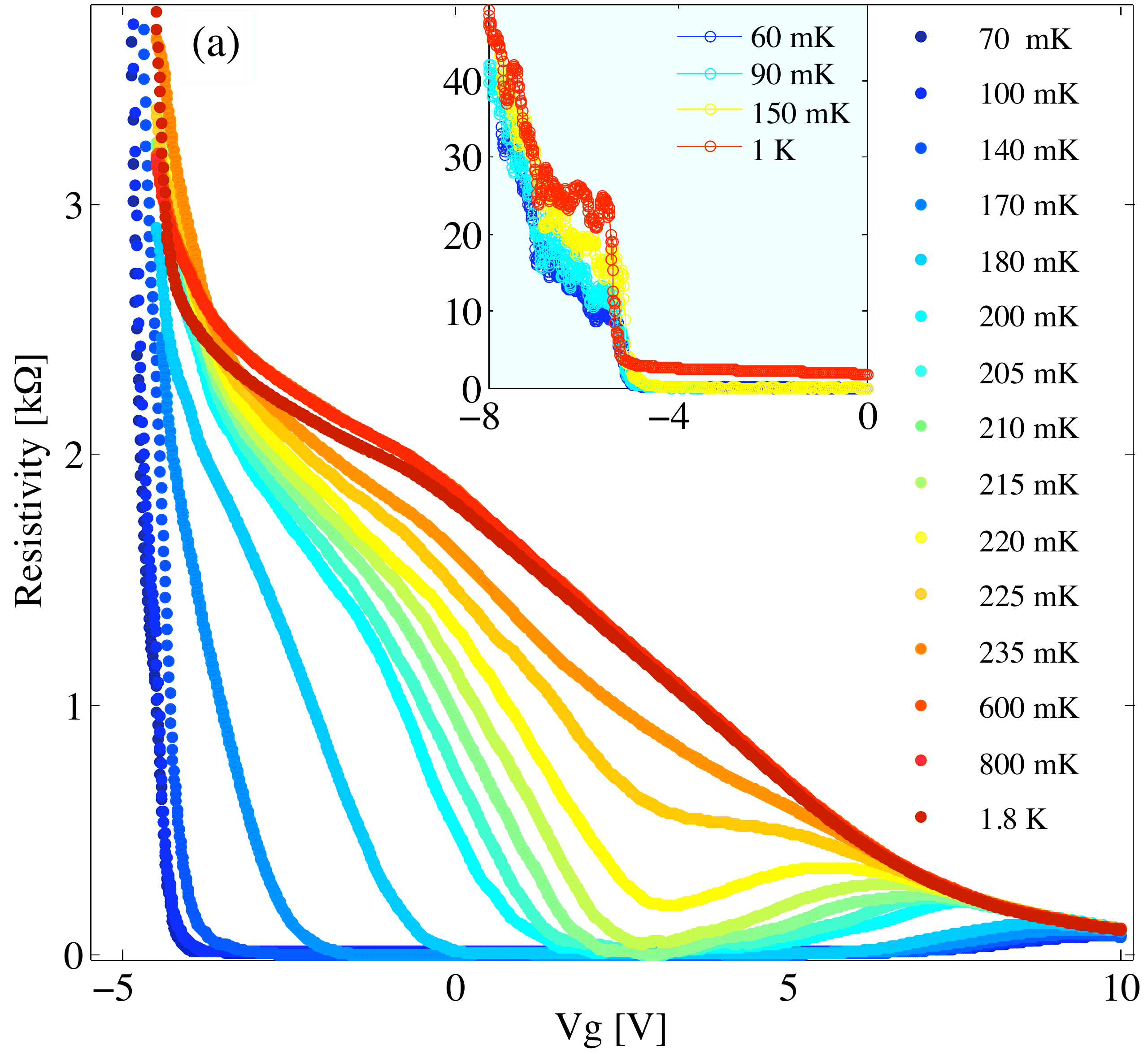}}
  \subfigure{\includegraphics[width=1\hsize]{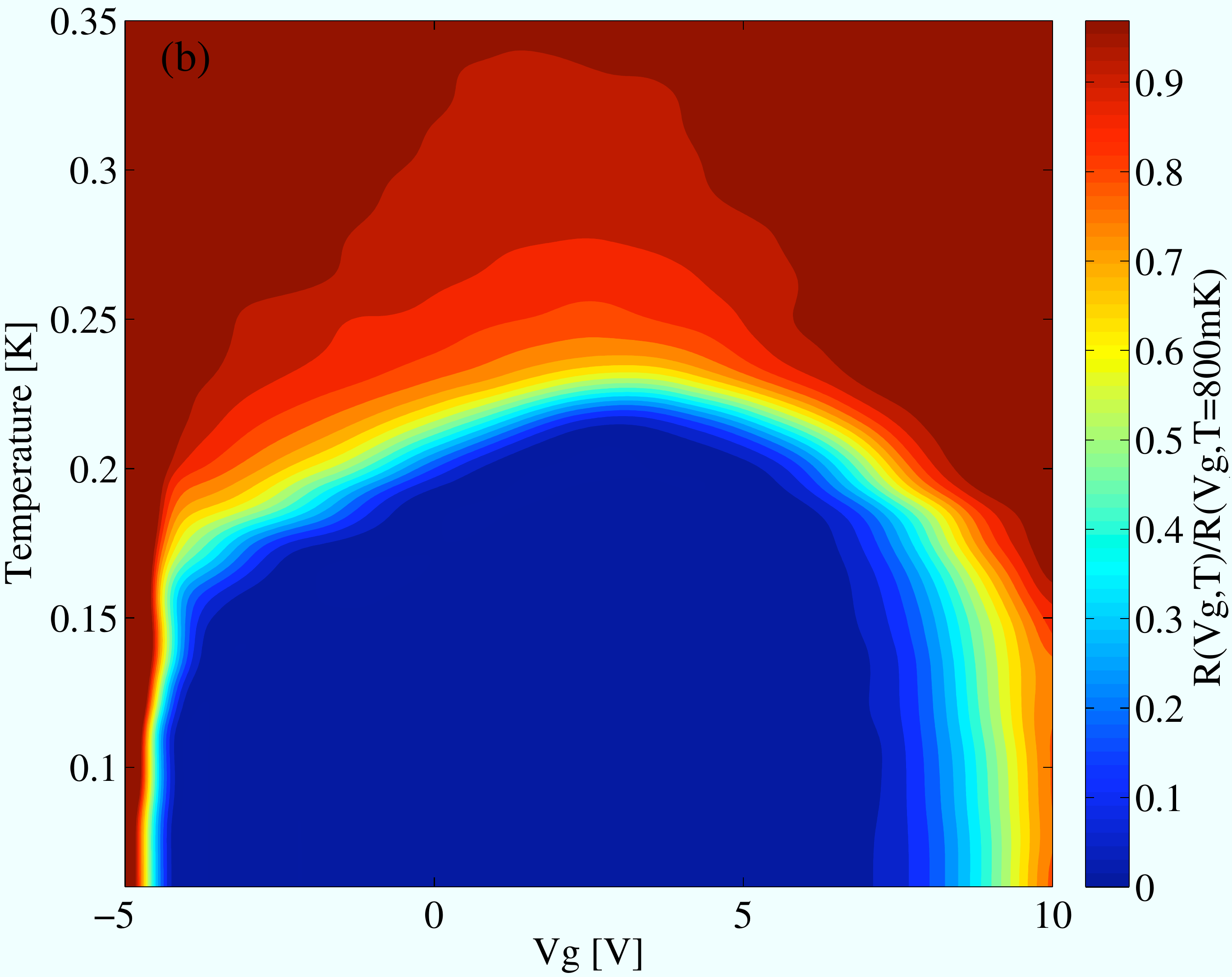}}
\caption {(a) The resistance per square R is plotted as a function of gate bias V$_g$ for various temperatures. Data is taken by scanning V$_g$ during a slow cool down to base temperature. Right Inset: additional data for lower V$_g$ values and higher temperatures. (b) Normalized resistance plotted as a function of gate bias and temperature. Different colors represent different fractions of the resistance at T=800 mK (above T$_c$) at the same value of V$_g$. \label{Tc_R}}
\end{center}
\end{figure}

\par
Most experiments exploring the phase diagram use large samples in back-gate configuration. This set up is prone to sample inhomogeneities and to the electric field dependence of the dielectric constant of the \STO~ substrate \cite{ChristenDielectricSTO1994}. In this paper we utilize our recently proposed geometry for back-gating \cite{rakhmilevitch2013anomalous} in which the large carrier concentration can be tuned using only a few volts as opposed to the hundreds of volts used in a standard back-gate geometry \cite{caviglia2008electric} minimizing the effects of the nonlinear response of \STO~ to electric field and spatial inhomogeneities. Our geometry allows us to carefully explore the evolution of the resistivity and the Hall signals across the critical points in the phase diagram. We find clear evidence that a second mobile band appears exactly where maximum T$_c$ and H$_c$ are observed and the \MR~ changes sign.
\par

We grow six epitaxial layers of  \LAO~ on atomically smooth \STO~substrate in standard conditions \cite{shalom2009anisotropic}. In the first step two unit cells of \LAO~ are deposited. Then a Hall bar of $9\times3$ $\mu m^{2}$ is patterned using electron beam lithography followed by deposition of a 40 nm thick amorphous BaTiO$_3$ layer and lift-off to define the conducting channel where an additional layer of 4 unit cells \LAO~ is epitaxially grown in the final step. For a schematic drawing of the device see figure \ref{device}. The design aims to minimize screening from the contact pads and leads, thus enhancing the gate bias effectivity. The strong response to back-gating is possible for dielectric substrates with high permittivity (high $\epsilon$) when the distance $d$ between the gate electrode and the tunable surface (or interface) becomes much larger than the width of the mesoscopic conducting channel $\ell$. This results in a capacitance per area of the order of $5\times 10^{12}$ [electrons$\times$cm$^{-2}$Volt$^{-1}$] for $d=0.5$ mm, $\ell$=3 $\mu$m, an order of magnitude larger than for standard planar capacitor geometry \cite{rakhmilevitch2013anomalous}.
\par
Gold gate electrodes are evaporated to cover the back of the substrate. The leakage current is unmeasurably small ($<$1pA). Measurements are performed while the sample is immersed in a diluted He$_3$/He$_4$ mixture with a base temperature of 60 mK. During cool down, the gate bias was scanned back and forth while recording the temperature and measuring the sheet (per-square) resistance, denoted as R. The gate scan is reversible within a 2 mV resolution, as long as the maximal gate bias (10 V) is not exceeded.
\par
Figure \ref{Tc_R}(a) presents the sheet resistance R versus gate voltage V$_g$ for several selected temperatures. For T=800 mK (upper red curve), decreasing V$_g$ from 10 V to -4 V results in a variation of R from $\approx$100 Ohms to $\approx$3 kOhms. Decreasing V$_g$ further, results in a sharp increase to R$\approx$25 kOhms (see inset).
For T=210 mK (green curve, arrow) R vanishes for V$_g$=3 V as the mesoscopic conducting channel becomes superconducting. R for the entire temperature, V$_g$ phase diagram is shown in figure \ref{Tc_R}(b). Here the resistance is normalized by its value at T=800 mK for the same gate voltage. Defining T$c$ as the temperature for which R(T$_c$) = R(T=800 mK)/2, the green color contour represents T$_c$ as a function of V$_g$. Maximum T$_c$ appears for V$_g=2.6\pm0.3$V. The entire superconducting dome is revealed by varying V$_g$ by merely 15 V thus demonstrating the performance of our gate design.
\par
The shape of the superconducting dome is somewhat different compared to what has been previously reported \cite{caviglia2008electric}. In particular, at the base temperature the SC transition as function of V$_g$ is very sharp. Upon increasing carrier density (over-doped region), the equi-resistance lines are almost perpendicular to the gate voltage axes.
This means that at low temperatures the transition becomes temperature independent, both in the over-doped and in the under-doped regimes. For the under-doped region this behavior persists into the normal state and will be discussed later.

\par
Caviglia \etal~ have shown that T$_c$ can be tuned from zero into the over-doped regime \cite{caviglia2008electric}. According to their capacitance and Hall measurements, the corresponding electron density modulation was $5\times10^{13}cm^{-2}$. Our previous study of high magnetic field Hall signal for large scale samples found a similar density modulation in the over-doped regime \cite{shalom2010tuning}. Such large density modulation is possible for a back gate geometry using \STO~as a dielectric material, which can withstand relatively high voltages. For a planar capacitor with thickness of $d$=0.5 mm and a low temperature dielectric ratio $k=\frac{\epsilon}{\epsilon_0}\approx 24000$ the capacitance per unit area, i.e. the 2DEG density modulation per bias voltage, is: $C_A=\frac{C}{A}=\frac{\epsilon}{ed} \simeq 2\times10^{11} cm^{-2}Volt^{-1}$. Taking into account the decreasing dielectric constant of \STO~as a function of the electric field \cite{dec1999scaling, ChristenDielectricSTO1994}, gate voltage as high as $\approx\pm$300 V are required for such modulation as has been confirmed experimentally \cite{caviglia2008electric}.
\par
The planar capacitor limit is $d<<\sqrt{A}$ with $d$ the distance between plates of area $A$. We utilize the opposite limit: $d>>\ell$, where $\ell$ is the width of the mesoscopic conducting channel (see figure \ref{Tc_R}(a), left inset). Usually, dielectric barriers are made as thin as possible - on the nanometer scale, making this limit irrelevant. However, in our design, the large thickness of the substrate makes the limit $d>>\ell$ feasible. For this unique geometry, the mutual capacitance between the mesoscopic channel and the gate decreases with $\ell$ but eventually saturates, governed only by $d$. Therefore, $C_A$ is enhanced becoming proportional to 1/$\ell$, before reaching the quantum capacitance limit. For the latter limit of a very small size: $\ell < \ell_{TF}=\epsilon/\nu$, ($\nu$ is the 2D density of states and $\ell_{TF}$ is the Thomas Fermi screening length) the capacitance is dominant by the quantum capacitance: $C_A\propto \frac{1}{\ell_{TF}}$, which may be the case for $\ell <$ 3 $\mu$m \cite{rakhmilevitch2013anomalous}.

In addition, for large charge density modulations $\epsilon$ is suppressed by the strong electric field \cite{dec1999scaling, ChristenDielectricSTO1994}, which leads to a substantial decrease in the geometrical capacitance at high gate biases. In the planar geometry this suppression extends through the entire thickness of the dielectric material, whereas in our device this effect takes place only in the vicinity of the mesoscopic channel. We estimated the decrease in $\frac{\delta n}{\delta V}$  due to such effect and found it to be approximately half that of the planar geometry for $\delta n=2\times10^{13}cm^{-2}$, roughly corresponding to a 5 V bias variation. This effect further enhances the gate efficiency.

\begin{figure}
 \begin{center}
    \subfigure{\includegraphics[width=1\hsize]{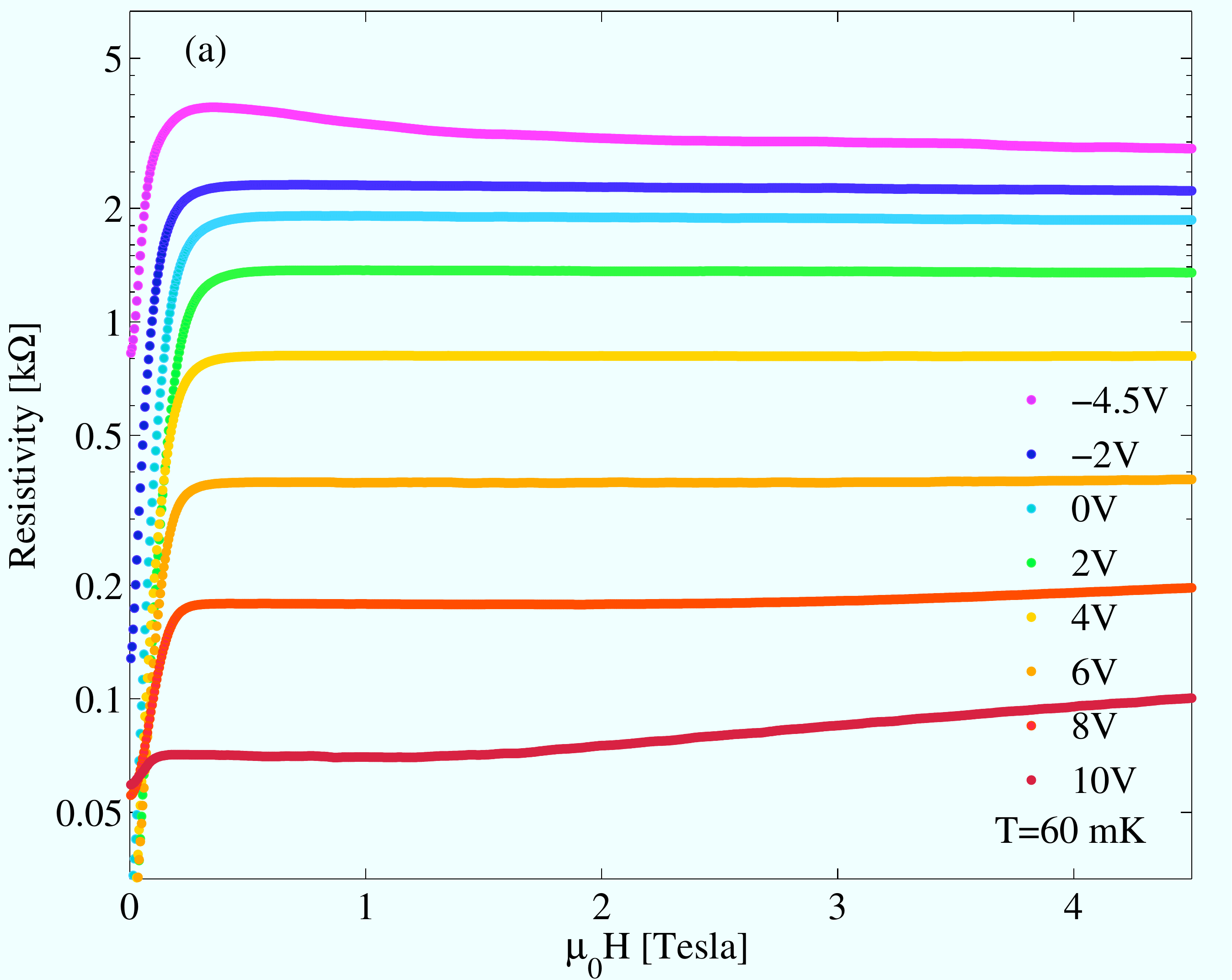}}
   \subfigure{\includegraphics[width=1\hsize]{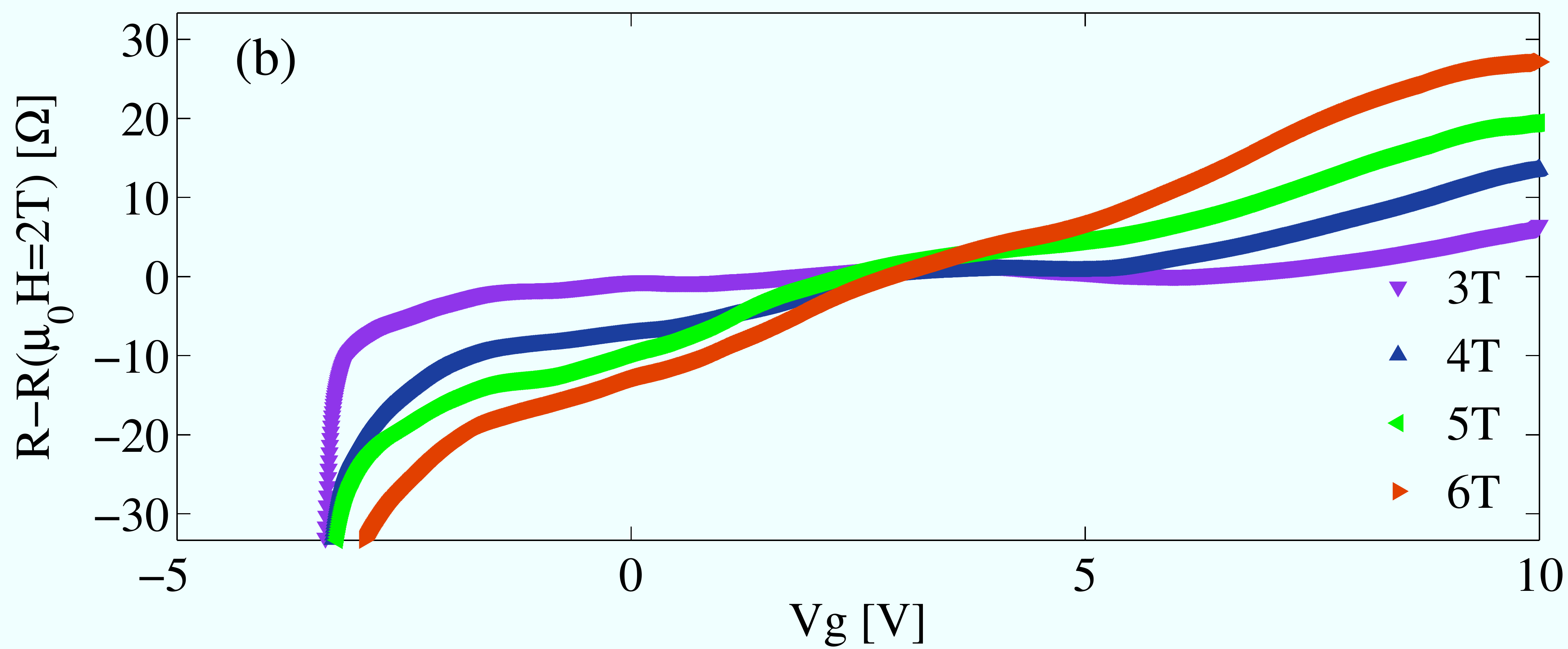}}
\caption {(a) The resistance per square, R is plotted on a logarithmic scale as a function of magnetic field at T=60 mK for various fixed values of gate bias V$_g$.(b) The magnetoresistance R(V$_g$)-R(V$_g$, $\mu_0$H=2 T) at various magnetic fields is plotted as a function of V$_g$. For different fixed magnetic fields: 6,5,4,3 T. We note a clear crossover from positive to negative magnetoresistance at 2.8$\pm0.45$ V.\label{Rxx_vsH}}
 \end{center}
\end{figure}
\begin{figure}
\includegraphics[width=1\hsize]{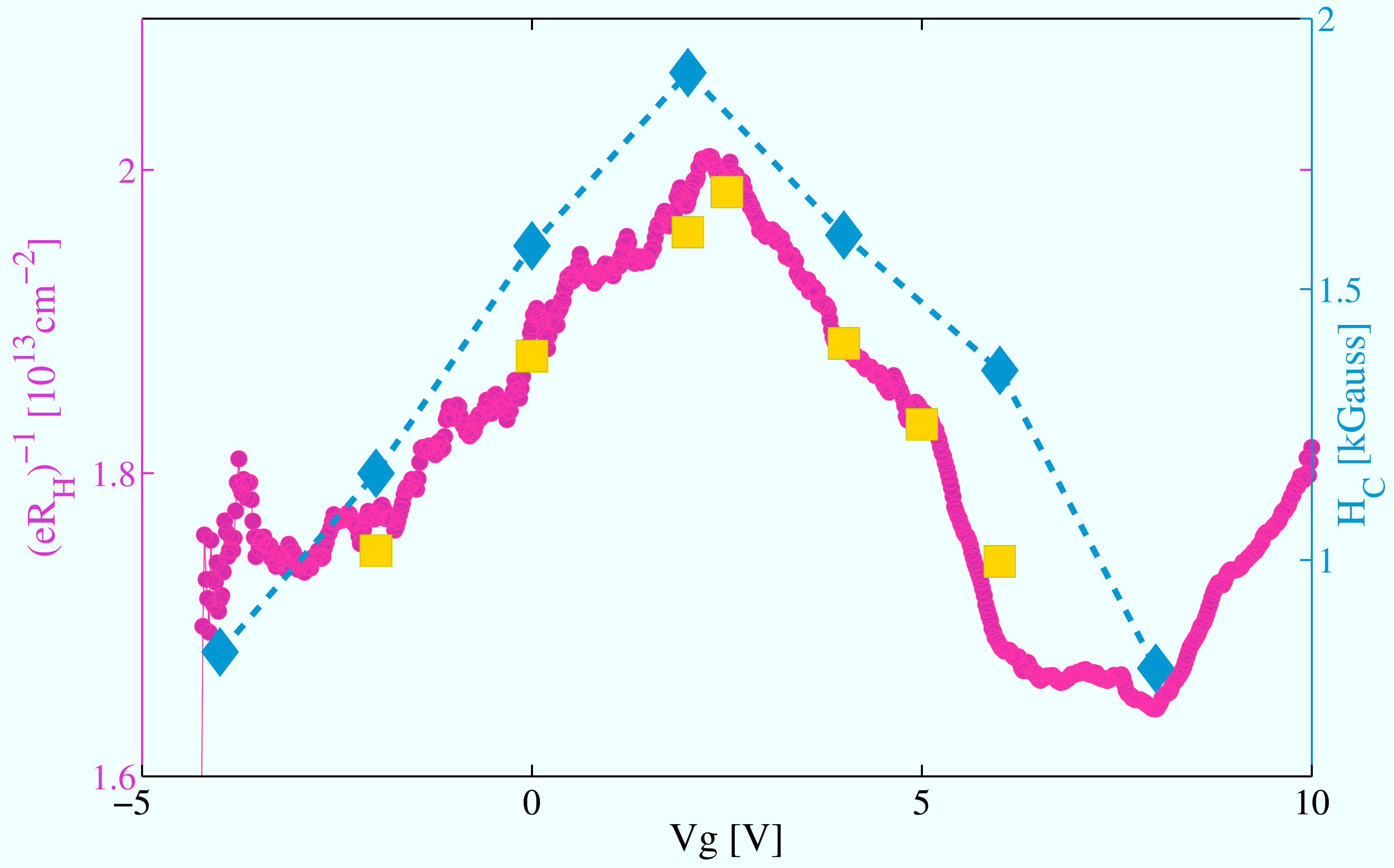}
\caption {Left axes: The inverse of the Hall slope is plotted as a function of the gate bias for a fixed magnetic field of 3 T (purple circles). The Hall coefficient is also measured while scanning the magnetic field from zero to 3 T for various fixed gates and fitting a linear curve (yellow squares). Right axes: Superconductivity critical magnetic field inferred from H$_{c}$ = H[R=1/2R(T=800 mK)] plotted for different fixed gates at T=60 mK (blue diamonds). \label{Density_Hc_IC_YY}}
\end{figure}
\par
The fine tunability of our device allows us to study in detail the dependence of magneto-transport properties and superconductivity on gate bias. Figure \ref{Rxx_vsH}(a) presents the longitudinal resistivity as function of magnetic field for several fixed gate biases. Superconductivity is suppressed at a critical magnetic field H$_c$. At magnetic fields higher than H$_c$ the MR depends on gate voltage. For example it is positive for V$_g$=10 V, and negative for V$_g$=-4.5 V. The detailed gate dependence of the MR at high magnetic fields is studied in figure \ref{Rxx_vsH}(b). R(H,V$_g$)-R(H=2T,V$_g$) is presented for various fixed magnetic fields and continuous gate scans. We choose R(H=2T) as the reference point to avoid non monotonic effects such as superconductivity and localization/anti-localization \cite{rakhmilevitch2010phase}. A distinct crossover from positive to negative MR occurs at V$_g$=2.8$\pm0.45$ Volt. Similar crossover was shown by Joshua \etal~\cite{joshua2012universal}.
\par
In figure \ref{Density_Hc_IC_YY} the inverse Hall coefficient $(eR_H)^{-1}$ in units of carrier density is plotted as a function of V$_g$ measured at base temperature. At the low temperature range under study $(eR_H)^{-1}$ is practically temperature independent. Purple circles, represent transverse voltage measured for $\mu_0$H = $\pm$3 T after antisymmetrizing the data. Yellow squares are data taken with a fixed V$_g$ while scanning the field from -3T to +3T and fitting a linear curve up to 3 T to the antisymmetrical part. The data taken using the two methods coincide strengthening the reliability of our measurements. A Clear non-monotonic signal is observed with a maximum at V$_g$=2.45$\pm0.3$ Volt. We also plot H$_c$ (Blue diamonds) defined using R(H$_c$)=1/2 R(T=800mK). H$_c$, T$_c$ and (eR$_H)^{-1}$ are all non-monotonic with a sharp maximum at the same density. At this density the MR changes sign. Both normal state properties (MR and (eR$_H)^{-1}$) and superconducting properties (H$_c$, T$_c$) point to the same gate voltage within error: V$_g$=2.6$\pm0.2$ Volt.
\par
We note that for the underdoped region, the sharp superconducting transition is accompanied by a sharp change in the normal state resistance (figure \ref{Tc_R}(a), right inset). Therefore, it may not characterize a generic superconductor-to-insulator transition and can originate from an extrinsic mechanism changing the normal state properties.
\par
The data presented here for such amplified charge-density modulation in a mesoscopic channel are unique in terms of the broad density range scanned in a single device, the sharp phase transitions and the non-monotonic Hall. For planar capacitor geometry the strong dependence of the dielectric constant on electric field results in a broadening of the phase transitions. This spurious effect is weaker in our design. However, it is not enough to explain the sharp transitions observed in figure \ref{Tc_R}(b). Although the geometry used is in principle more sensitive to fringing electric fields the very sharp transitions observed can only be explained by homogenous 2DEG characteristics relative to macroscopic samples.
\par
It has recently been shown that the tetragonal domains of the \STO~ can effect the local potential and conductance at the \STO/\LAO~ interface \cite{kalisky2013locally, honig2013local}. We believe that the small gate voltage range used helps us to avoid effects caused by strong domain wall motion induced by the back gate. The reproducibility of our results and absence of hysteresis despite many successive cool down and gate voltage cycles support this view.
\par
The sharp nonmonotonic behavior of (eR$_H)^{-1}$ (figure \ref{Density_Hc_IC_YY}) can, in principle, be explained by a simple two-band model. At low gate voltages only the low mobility band is present. The maximum in (eR$_H)^{-1}$ appears at the onset of the second band population, which according to this simple model has a higher mobility. The data is in agreement with a high mobility which is about three times as large as the mobility of the lower band. These results are consistent with our previous measurements. \cite{shalom2010shubnikov, lerer2011low, rakhmilevitch2013anomalous}.
\par
According to the simple two-band scenario suggested above, in the underdoped (low carrier) regime where only a single band is populated the low field Hall voltage slope should be a direct measurement of the carrier concentration. Using this assumption we obtain a changes in carrier concentration from $\approx$ 1.7 $\times10^{13}$ to 2 $\times10^{13} cm^{-2}$ when changing the gate voltage from V$_g$ = -4 to V$_g$=3 Volts respectively.
\par
However, if one assumes that changes in T$_c$ are driven solely by the carrier concentration, namely, that T$_c$ is a well defined function of n, one finds that the observed modulation in carrier concentration is very low compared to the modulation needed for the same changes in T$_c$ found in larger samples. For example: for similar variation in T$_c$ reported in Refs.\onlinecite{caviglia2008electric, PhysRevLett.104.126802} an order of magnitude larger density modulation was found according to the obtained C$_A$ and by the Hall signal. We can therefore conclude that while the two band picture is good enough to explain the change in the behavior of $R_H$ it is not sufficient to explain the different behavior of mesoscopic samples and large samples; in particular, the order of magnitude difference in variation in the carrier concentration for a known modulation of T$_c$.
\par
Hosoda \etal~ \cite{:/content/aip/journal/apl/103/10/10.1063/1.4820449} showed that mobility depends on gate configuration: top gate increase mobility upon depletion as opposed to back gate. This may be a result of pushing the conducting electrons away from (in top gate configuration) or towards (back gate) the interface upon depletion. Our unconventional gating geometry may result in a different distance of the carriers from the interface, which results in a different mobility in comparison with the planar geometry. This effect may partially explain the different behaviors of the Hall slope for these geometries. Another effect that may be important in mesoscopic samples is a more dominant role of the quantum capacitance, this effect is overwhelmed by the geometrical capacitance per unit area for the large samples measured so far. We speculate that a more dominant role of the quantum capacitance (relative to the geometrical one) may explain the difference between our mesoscopic conducting channel and large samples.

\par
In summary, we found a direct link between superconductivity and multiple band occupation in \STO/\LAO~interfaces. Maximum T$_c$ appears exactly at the same gate bias where a maximum in $(eR_H)^{-1}$ is observed. The geometry used to gate and thus probe the electronic properties of the interface as a function of carrier concentration can be implemented to other materials such as graphene and topological insulators assembled on top of a high $\epsilon$ dielectric substrate.
\begin{acknowledgments}
This research was partially supported by the Israel Science Foundation
(ISF) under grant 569/13  and the Ministry of Science and Technology under contract 3-8667 and the Bi-National Science Foundation (BSF) under grant 2010140.
\end{acknowledgments}
\bibliography{mybib}
\end{document}